\documentclass[sigconf]{acmart}
\acmConference[EASE 2026]{The 30th International Conference on Evaluation and Assessment in Software Engineering}{June 9--12, 2026}{Glasgow, Scotland, United Kingdom}
\acmBooktitle{Proceedings of the 30th International Conference on Evaluation and Assessment in Software Engineering (EASE 2026), June 9--12, 2026, Glasgow, Scotland, United Kingdom}
\copyrightyear{2026}
\acmYear{2026}
\setcopyright{acmlicensed}
\makeatletter
\@ACM@balancefalse
\makeatother

\usepackage{amsmath,amsfonts}
\usepackage{graphicx}
\usepackage{xcolor}
\usepackage{adjustbox}
\usepackage{multirow}
\usepackage{enumitem}
\usepackage{booktabs}

\setlist[itemize]{leftmargin=*,nosep}
\setlist[enumerate]{leftmargin=*,nosep}
\begin{document}

\title{Multifaceted Hero Developers and Bug-Fixing Outcomes Across Severity}

\author{Amit Kumar}
\email{amitchandramunityagi@gmail.com}
\affiliation{%
  \institution{IIIT Allahabad}
  \country{India}
}

\author{Mahen Gandhi}
\email{mahengandhi19@gmail.com}
\affiliation{%
  \institution{IIIT Hyderabad}
  \country{India}
}

\author{Meher Bhardwaj}
\email{bhardwajmeher01@gmail.com}
\affiliation{%
  \institution{IIIT Manipur}
  \country{India}
}

\author{Hrishikesh Ethari}
\email{hrishikeshethari@gmail.com}
\affiliation{%
  \institution{IIIT Manipur}
  \country{India}
}

\author{Sonali Agarwal}
\email{sonali@iiita.ac.in}
\affiliation{%
  \institution{IIIT Allahabad}
  \country{India}
}

\begin{abstract}
Open-source projects often rely on a small group of highly active contributors known as \emph{hero developers}. Prior work shows that hero developers are common in many OSS and enterprise projects, yet who qualifies as a hero depends heavily on the chosen contribution metric. Code-based metrics identify implementation-focused developers, whereas discussion-based metrics highlight coordination and communication; these metrics capture distinct \emph{facets} of contribution. We conducted a measurement-sensitive study of multifaceted heroism across 77 Apache Software Foundation projects using three technical measures (commit count, distinct files touched, churn) and two social measures (issue-comment count, number of distinct issues commented on). We examined hero prevalence, overlap among hero sets, and severity-wise bug-fixing outcomes via fix and reopen rates. Results show that hero projects are common under all measures, but identified heroes differ substantially across facets. The pooled Jaccard overlap between technical and social hero sets is only 0.10. Cross-facet asymmetry is evident: 71.4\% of technical heroes exhibit strong social activity, while only 24.2\% of social heroes show strong technical activity. Fix-rate and reopen-rate differences are modest, yet hero-category rankings vary across severity levels and outcome measures. These findings indicate that heroism is not a single, metric-independent role. A multifaceted perspective offers a more reliable understanding of key contributors and better supports developer prioritisation and severity-aware bug assignment.
\end{abstract}

\keywords{hero developers, hero projects, technical contribution, social contribution, multifaceted heroism, robustness, measurement sensitivity}
\ccsdesc[500]{Software and its engineering~Software creation and management}

\maketitle

\section{Introduction}
Software development effort in open-source software (OSS) projects is rarely distributed evenly. A small group of highly active developers often accounts for a disproportionately large share of the work. Prior studies define such projects as \emph{hero projects} when 20\% or fewer developers perform at least 80\% of the contributions, and refer to these highly active contributors as \emph{hero developers}~\cite{robles2006contributor,ricca2010heroes,agrawal2018we}.

Early research viewed heavy reliance on hero developers as a sustainability risk due to potential loss of key knowledge and project continuity~\cite{robles2006contributor,ricca2010heroes}. In contrast, recent studies show that hero projects are common in many OSS projects and are not necessarily linked to negative outcomes~\cite{agrawal2018we}. Moreover, Majumdar et al.~\cite{majumder2019software} found that hero developers may positively affect software quality by introducing fewer bugs than non-heroes. These findings suggest that heroism may reflect not only risk, but also concentrated expertise and critical contributions.

However, who is labelled a hero depends heavily on how contribution is measured. Developers may be identified as heroes through high commit counts, extensive file modifications, or large code churn (technical contributions), or through active issue discussion, bug triage, and coordination (social contributions). These represent distinct yet complementary facets of socio-technical development~\cite{herbsleb2014socio}.

Prior work has examined technical and social heroes separately, but two gaps remain: (1) it is unclear whether different measures identify the same developers as heroes, and (2) less is known about \emph{multifaceted} developers who are highly active in both technical and social dimensions.

To address these gaps, we study \emph{multifaceted heroism} in OSS. We analyse five contribution measures---three technical (commit count, distinct files touched, churn) and two social (issue-comment count, distinct issues commented on). We examine the prevalence of heroism, the overlap among hero sets defined by different measures, and the bug-fixing performance of various hero categories across bug severity levels, using both fix rate and reopen rate.

We conduct an empirical study on 77 Apache Software Foundation projects from the SmartSHARK dataset. For each project, we identify hero developers using a commonly used 20/80 operationalisation and analyse how hero identification and bug-fixing effectiveness vary across contribution measures and hero categories. We also evaluate the robustness of results under alternative thresholds and attribution methods.

The study is guided by three research questions:

\begin{itemize}
    \item \textbf{RQ1:} How prevalent are hero projects and hero developers across technical and social measures, and how common is multifaceted heroism?
    \item \textbf{RQ2:} How consistent are hero classifications across different contribution measures and hero categories?
    \item \textbf{RQ3:} How do hero categories compare in bug-fixing outcomes (fix rate and reopen rate) across severity levels, and how robust are these findings under alternative hero definitions and attribution choices?
\end{itemize}

Our results indicate that hero projects are common across all measures, but the specific developers identified as heroes differ substantially. Technical and social hero sets show low overlap, revealing that implementation and coordination concentration are not equivalent. We observe an asymmetric pattern: technical heroes often exhibit strong social activity, while social heroes less frequently show strong technical activity. In bug fixing, absolute performance differences among hero categories are modest, but their relative rankings vary by severity level and outcome measure.

The main contributions of this paper are:
\begin{itemize}
    \item A measurement-sensitive analysis of hero developer identification across technical and social contribution facets.
    \item An empirical study of multifaceted heroism, focusing on developers active in both technical and social dimensions.
    \item Quantification of overlap among hero sets, demonstrating strong dependence on the chosen contribution measure.
    \item Evaluation of severity-wise bug-fixing performance using both fix rate and reopen rate.
\end{itemize}

\vspace{-0.35\baselineskip}
\section{Related Work}
Uneven contribution is a recurring observation in FLOSS research. Early work showed that a small core often performs much of the work and that this core can evolve over time~\cite{robles2009evolution,robles2006contributor}. Related truck/bus factor studies model the risk of dependence on a small set of knowledgeable developers~\cite{ricca2010heroes,jabrayilzade2022busfactor}. Hero studies focus on a related but distinct issue: contribution concentration and its association with outcomes. Empirical evidence is mixed. Agrawal et al.~\cite{agrawal2018we} report that hero projects are common and not necessarily worse on process outcomes, while socio-technical work argues that coordination shaped by technical dependencies can influence quality~\cite{herbsleb2014socio,majumder2019software}.

A key methodological issue is that ``hero'' or ``key developer'' is induced by a measurement rule. Some studies use VCS activity such as commits~\cite{agrawal2018we}; others use code-centric contribution measures~\cite{tsikerdekis2018persistent}, artifact traceability graphs~\cite{ccetin2020identifying}, or privileged events in issue and pull-request workflows~\cite{bock2023automatic}. Industry and open-source work increasingly uses multi-dimensional contribution models~\cite{sun2024realhero,li2025measuring,li2024memento}. Such aggregation can be useful, but it may hide which facet drives a result. We therefore keep technical and social facets separate, quantify overlap among induced hero sets, and relate the resulting categories to bug-fixing outcomes grounded in prior defect-fixing and bug-reopen research~\cite{ghapanchi2011effectiveness,zimmermann2012characterizing,de2022studying} using SmartSHARK artifacts~\cite{trautsch2021msr}.

\section{Methodology}
For each project, we compute five contribution measures (three technical, two social) and identify whether the project is heroic under each measure using the commonly used 20/80 operationalisation~\cite{ricca2010heroes,agrawal2018we}. We then derive multifaceted categories using a cross-facet ``strong'' threshold (default median) and report results at two granularities: (i) 23 fine-grained categories (five base + 18 pair-derived), and (ii) five aggregated hero types for primary reporting.

\subsection{Dataset Overview}
We use SmartSHARK~2.1, a repository-mining corpus integrating version-control and issue-tracking artifacts for Apache projects~\cite{trautsch2021msr}. The dataset contains 77 projects, 366{,}322 commits, 163{,}057 issues, and 772{,}883 issue comments~\cite{trautsch2021msr}. It covers commit author timestamps from 1999-11-09 to 2021-05-17 and issue activity up to July 2021.

\par \textbf{Identity linkage (cross-artifact developers).}
We treat the SmartSHARK developer entity as the unit of analysis and link commit authors, issue commenters, and assignees using the identity mapping provided by SmartSHARK. When cross-artifact links are missing, identities remain distinct (i.e., we do not perform additional manual/heuristic merging).

\par For RQ3 we require at least 80\% of bug issues in a project to have a non-missing priority value. Two projects fail this threshold and are excluded from RQ3, leaving 75 projects eligible for outcome analyses.

\subsection{Contribution Measures}
\label{sec:measures}
We define five contribution measures at the project level. For a developer $x$ in project $p$:
\begin{itemize}
\item \textbf{Technical measures}
  \begin{itemize}
  \item \textbf{T1 (Commit count):} number of commits authored by $x$ in $p$.
  \item \textbf{T2 (Distinct files touched):} number of distinct files touched by commits authored by $x$ in $p$.
  \item \textbf{T3 (Churn):} total churn by $x$ in $p$, computed as $\sum(\textit{lines\_added}+\textit{lines\_deleted})$ over file actions in commits authored by $x$.
  \end{itemize}
\item \textbf{Social measures}
  \begin{itemize}
  \item \textbf{S1 (Comment count):} number of issue comments authored by $x$ in $p$.
  \item \textbf{S2 (Distinct issues commented on):} number of distinct issues in $p$ on which $x$ authored at least one comment.
  \end{itemize}
\end{itemize}

\par Technical measures use commit \emph{authors} (not committers) to reduce credit assignment to integrators. We interpret churn as \emph{change magnitude}, not ``lines written.'' Our social facet is operationalised via issue-discussion participation in the issue tracking system; to maintain comparability across projects and reduce workflow heterogeneity, we restrict social measures to issue comments.

\subsection{Hero Identification and Category Construction}
\label{sec:heroes}
We define heroism per project and per measure following the Pareto-style operationalisation common in hero literature~\cite{ricca2010heroes,agrawal2018we}. Let $C_{p,m}(x)$ denote the contribution of developer $x$ in project $p$ under measure $m \in \{T1,T2,T3,S1,S2\}$. Let $D_{p,m}=\{x \mid C_{p,m}(x) > 0\}$ and $n_{p,m}=|D_{p,m}|$.

\subsubsection{Hero projects and hero developers (base measures)}
We rank developers in $D_{p,m}$ by $C_{p,m}(x)$ and select the top
$k_{p,m}=\max(1,\lceil \alpha \cdot n_{p,m}\rceil)$ developers, where $\alpha=0.20$ by default. Ties are broken deterministically by developer identifier (ascending). Let $\mathrm{Top}_{p,m}(\alpha)$ be this set. Project $p$ is heroic under measure $m$ if
\[
  \frac{\sum_{x \in \mathrm{Top}_{p,m}(\alpha)} C_{p,m}(x)}{\sum_{x \in D_{p,m}} C_{p,m}(x)} \ge \beta,
\]
with $\beta=0.80$ by default. If the condition holds, the hero developer set is $H_{p,m}=\mathrm{Top}_{p,m}(\alpha)$; otherwise $H_{p,m}=\emptyset$.

\subsubsection{Strong activity in the other facet}
To construct multifaceted categories, we require strong activity on the other facet within the same project. For each project $p$ and measure $m$, let $Q_{p,m}(q)$ denote the $q$-quantile of $\{C_{p,m}(x)\mid x\in D_{p,m}\}$. Unless stated otherwise, we use $q=0.50$ (median) as a minimal ``strong'' threshold. We assess $q{=}0.75$ in sensitivity analyses.

\subsubsection{Pair-derived categories (18)}
For each technical--social pair $(T_i,S_j)$ with $i\in\{1,2,3\}$ and $j\in\{1,2\}$:
\begin{align*}
  \mathrm{TS}_{p,i,j}(q) & =
  \{x \mid x \in H_{p,T_i} \ \wedge\ C_{p,S_j}(x) \ge Q_{p,S_j}(q)\}, \\
  \mathrm{ST}_{p,i,j}(q) & =
  \{x \mid x \in H_{p,S_j} \ \wedge\ C_{p,T_i}(x) \ge Q_{p,T_i}(q)\}, \\
  \mathrm{SH}_{p,i,j}    & = H_{p,T_i} \cap H_{p,S_j}.
\end{align*}

\subsubsection{Aggregated hero types}
For primary reporting we aggregate within a project:
\[
\begin{adjustbox}{max width=\columnwidth,center}
$\begin{gathered}
H^{\textit{tech}}_p = \bigcup_{i=1}^{3} H_{p,T_i}, \qquad
H^{\textit{soc}}_p  = \bigcup_{j=1}^{2} H_{p,S_j}, \\
H^{\textit{ts}}_p   = \bigcup_{i=1}^{3}\bigcup_{j=1}^{2} \mathrm{TS}_{p,i,j}(q), \qquad
H^{\textit{st}}_p   = \bigcup_{i=1}^{3}\bigcup_{j=1}^{2} \mathrm{ST}_{p,i,j}(q), \qquad
H^{\textit{sup}}_p  = \bigcup_{i=1}^{3}\bigcup_{j=1}^{2} \mathrm{SH}_{p,i,j}.
\end{gathered}$
\end{adjustbox}
\]
Superheroes are therefore developers who satisfy at least one technical and at least one social hero definition within the same project.

\subsection{Setup to Address the Research Questions}
\label{sec:rqsetup}

\subsubsection{RQ1: Prevalence}
For each base measure $m$, hero-project prevalence is the fraction of projects with $H_{p,m}\neq\emptyset$. We report prevalence for aggregated hero types similarly. For multifaceted prevalence we compute conditional proportions such as $\frac{|\mathrm{TS}_{p,i,j}(q)|}{|H_{p,T_i}|}$ and $\frac{|\mathrm{ST}_{p,i,j}(q)|}{|H_{p,S_j}|}$ (when denominators are non-empty).

\subsubsection{RQ2: Overlap and stability}
We quantify overlap with the Jaccard coefficient $J(A,B)=|A\cap B|/|A\cup B|$, because hero identification is fundamentally a set-membership question. We report (i) global overlap by pooling project--developer memberships across projects, and (ii) per-project overlap summarised by median and IQR. For per-project summaries, we include every project with non-empty $A\cup B$; if only one set is non-empty in a project, that project contributes $J{=}0$. We additionally compute cross-facet Spearman correlations between technical and social contribution values per project to characterise monotone rank alignment under heavy-tailed contribution distributions.

\subsubsection{RQ3: Bug-fixing outcomes and ranking stability}
We focus on bug fixing because it is a consequential maintenance activity and because SmartSHARK provides linked bug reports, priorities, assignees, and fixing commits that enable severity-aware analysis without collecting additional data. We therefore restrict to issues labelled as \emph{bugs} and proxy severity using issue priority. We retain priorities in \emph{blocker}, \emph{critical}, \emph{major}, \emph{minor}, or \emph{trivial} (case-insensitive). We consider an issue \emph{fixed} if its history contains a resolution change to \emph{fixed} or its final resolution is \emph{fixed}. We mark an issue as \emph{reopened-after-fixed} if, after the first resolution-to-fixed event, any later status change sets the issue to \emph{reopened}; reopenings are binary per issue~\cite{zimmermann2012characterizing}.

\par We evaluate \emph{fix rate} and \emph{reopen rate}~\cite{ghapanchi2011effectiveness,zimmermann2012characterizing}. For a project $p$, severity $s$, and hero category $g$, let $n_{\textit{assigned}}$ be the number of bugs attributed to $g$ under a bug attribution policy, and $n_{\textit{fixed}}$ and $n_{\textit{reopened}}$ the counts of fixed and reopened-after-fixed bugs. We compute:
\[
  \textit{fix\_rate} = 100 \cdot \frac{n_{\textit{fixed}}}{n_{\textit{assigned}}}, \qquad
  \textit{reopen\_rate} = 100 \cdot \frac{n_{\textit{reopened}}}{n_{\textit{fixed}}}.
\]
Division-by-zero cases are treated as missing: if $n_{\textit{assigned}}{=}0$, fix rate is undefined; if $n_{\textit{fixed}}{=}0$, reopen rate is undefined.

\par Our primary attribution policy is \emph{assignee-based}: a bug is attributed to a category if at least one assignee belongs to that category in the corresponding project. As robustness we also consider \emph{assignee-at-fix-time} and a combined policy that attributes if either the assignee-at-fix-time belongs to the category or a linked fixing commit author belongs to it.

\par We summarise category performance with project-level medians because rates are heterogeneous across projects and ties are common. To assess whether severity changes the relative ordering of categories, we compare severity-wise rankings using Rank-Biased Overlap (RBO; $p=0.9$), which emphasises agreement near the top of short ranked lists. Rankings are computed on unrounded median rates. When equal medians occur, ties are broken deterministically using a fixed aggregated-type order (technical, social, sociotechnical, technosocial, super). We report upper-tail permutation-test p-values, computed exactly when the compared list depth is at most 8 and otherwise with 10{,}000 Monte Carlo permutations.

\subsection{Robustness and Sensitivity Analyses}
\label{sec:robustness}
We assess sensitivity to threshold choices inducing hero labels. In addition to the default $(\alpha,\beta,q)=(0.20,0.80,0.50)$, we vary $\alpha \in \{0.15,0.20,0.25,0.30\}$, $\beta \in \{0.70,0.80,0.85\}$, and $q \in \{0.50,0.75\}$. For RQ3, we assess robustness to bug attribution by comparing assignee-based results against assignee-at-fix-time and combined variants. Complete sweep outputs are provided in the replication package.

\begin{table}[!t]
\caption{Hero-project prevalence for base measures and aggregated hero types ($N{=}77$). Multifaceted rows show $q$-sensitivity at default $(\alpha,\beta)$.}
\label{tab:rq1-prevalence}
\centering
\footnotesize
\setlength{\tabcolsep}{2pt}
\begin{tabular*}{\columnwidth}{@{\extracolsep{\fill}}llrr@{}}
\toprule
\textbf{Definition} & \textbf{Measure / type} & \textbf{\# Hero projects} & \textbf{Prev. (\%)} \\
\midrule
\multirow{5}{*}{Base measures}
  & T1                               & 64 & 83.1 \\
  & T2                               & 64 & 83.1 \\
  & T3                               & 75 & 97.4 \\
  & S1                               & 71 & 92.2 \\
  & S2                               & 62 & 80.5 \\
\midrule
\multirow{5}{*}{Aggregated hero types}
  & Technical ($H^{\textit{tech}}$)          & 76 & 98.7 \\
  & Social ($H^{\textit{soc}}$)              & 71 & 92.2 \\
  & Sociotechnical ($H^{\textit{st}}$, $q{=}0.50$)       & 71 & 92.2 \\
  & Technosocial ($H^{\textit{ts}}$, $q{=}0.50$)         & 76 & 98.7 \\
  & Superhero ($H^{\textit{sup}}$)           & 70 & 90.9 \\
\midrule
\multirow{2}{*}{Multifaceted sensitivity}
  & Sociotechnical ($H^{\textit{st}}$, $q{=}0.75$)       & 71 & 92.2 \\
  & Technosocial ($H^{\textit{ts}}$, $q{=}0.75$)         & 76 & 98.7 \\
\bottomrule
\end{tabular*}
\raggedright\footnotesize\emph{Developer-level shift:} ST*/S* median 24.2\%\,$\rightarrow$\,18.2\% and rows 1,446\,$\rightarrow$\,1,026 (-29.0\%); TS*/T* median 71.4\%\,$\rightarrow$\,67.4\% and rows 951\,$\rightarrow$\,902 (-5.2\%).
\end{table}

\section{Results and Discussion}
\label{sec:results}

\subsection{RQ1: Prevalence of heroism and multifaceted heroism}
Table~\ref{tab:rq1-prevalence} reports hero-project prevalence under each base measure and under the aggregated hero types. It also compares the default cross-facet threshold ($q{=}0.50$) with the stricter threshold ($q{=}0.75$).

\subsubsection{Hero projects are common across measures}
Under the commonly used 20/80 rule, most projects are heroic regardless of the measure used. Churn-based heroism (T3) yields the highest base-measure prevalence (75/77 projects). Commit count (T1) and distinct files touched (T2) each identify 64/77 hero projects. Social measures show the same overall pattern: issue-comment count (S1) identifies 71/77 hero projects, and distinct issues commented on (S2) identifies 62/77. Thus, the presence of concentration is not an artifact of one metric.

\subsubsection{Cross-facet strength is common but asymmetric}
At the project level, multifaceted categories are also widespread: sociotechnical heroes appear in 71/77 projects, technosocial heroes in 76/77, and superheroes in 70/77. The developer-level picture is more informative. A median of 71.4\% of technical heroes are also technosocial, whereas only 24.2\% of social heroes are also sociotechnical. In other words, many high-activity implementers also participate visibly in issue discussions, but many highly active discussants are not correspondingly high-volume implementers.

A plausible explanation is role heterogeneity. Developers who change substantial amounts of code often need to coordinate fixes, clarify implementation choices, or respond to bug reports. Issue discussions, however, include several roles beyond implementation: triagers, reporters, users, and maintainers who guide work without authoring many commits. This interpretation is consistent with the low strict intersection: the median fraction of developers in the technical--social union who are heroes in both facets is 11.1\% (IQR 4.4--18.2\%).

Tightening the cross-facet threshold to $q{=}0.75$ leaves project-level prevalence unchanged but reduces the number of developers counted as multifaceted (Table~\ref{tab:rq1-prevalence}). This is an important distinction: many projects contain at least one multifaceted hero, but the size of the multifaceted group depends on the threshold.

\par \textbf{Answer to RQ1:} \textit{Hero projects are common across technical and social measures. Multifaceted heroism is also common, but asymmetric: technical heroes frequently show strong social activity, while social heroes less often show strong technical activity.}

\begin{figure}[!t]
\centering
\includegraphics[width=\columnwidth]{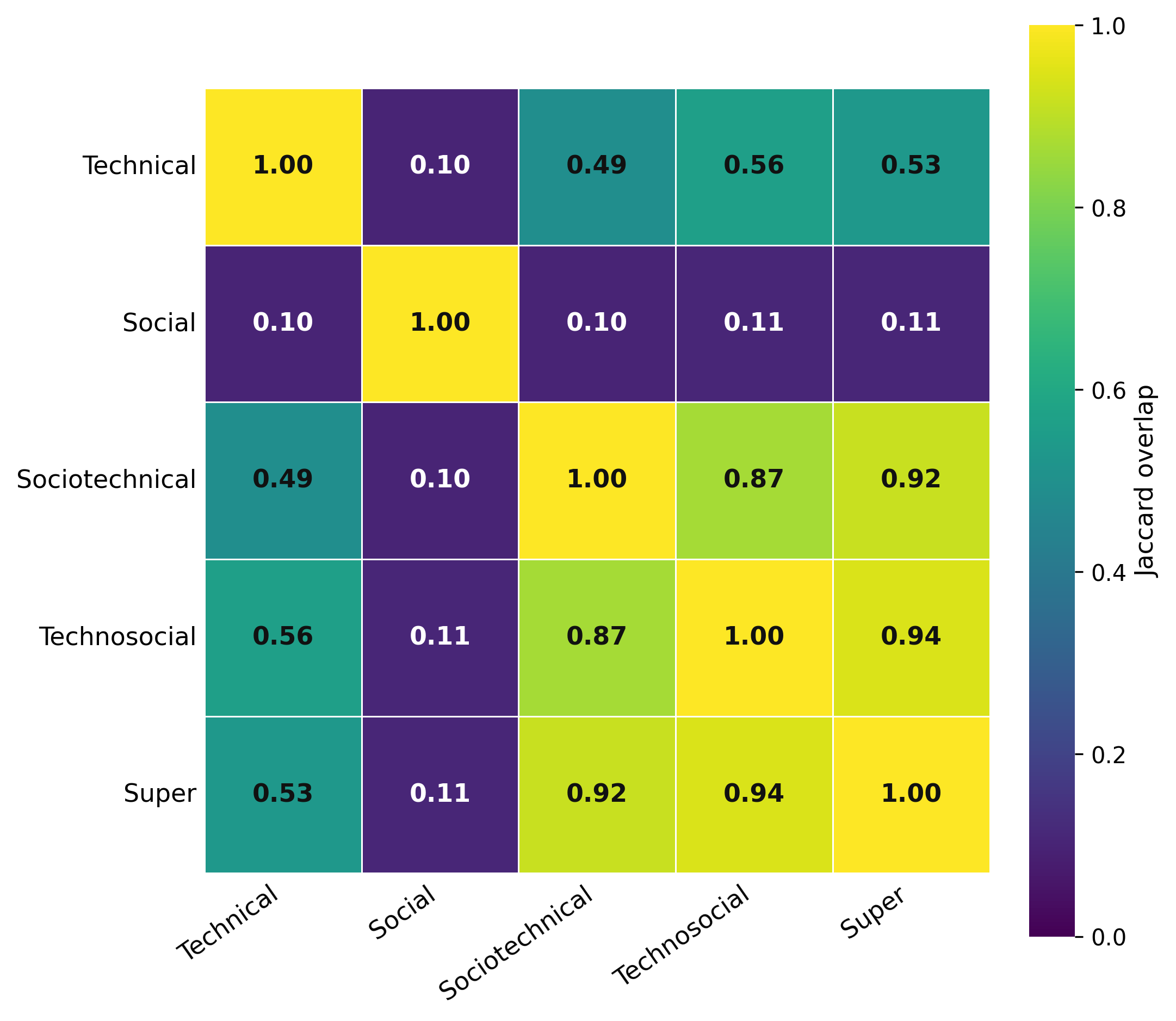}
\caption{Global pooled Jaccard overlap among aggregated hero types over project--developer memberships.}
\Description{Heatmap of pooled Jaccard overlap among technical, social, sociotechnical, technosocial, and superhero hero sets. The technical-social overlap is low, around 0.10, while overlaps among multifaceted types are high.}
\label{fig:rq2-heatmap}
\end{figure}

\subsection{RQ2: Consistency and overlap across measures and hero types}
\subsubsection{Within-facet overlap is moderate to high}
Technical measures tend to identify similar, though not identical, developers. Global Jaccard overlaps among T1, T2, and T3 range from 0.575 to 0.724; median per-project overlaps range from 0.586 to 0.714. The two social measures also agree substantially (S1 vs. S2: global $J{=}0.593$, median per-project $J{=}0.624$). This suggests that alternative measures within the same facet usually capture related forms of activity.

\subsubsection{Across-facet overlap is low}
The main disagreement appears across facets. Aggregated technical and social hero sets have a pooled Jaccard overlap of only 0.103 (Figure~\ref{fig:rq2-heatmap}), with a median per-project Jaccard of 0.103. This means that a developer identified as a technical hero is often not the same person identified as a social hero, and vice versa. A single hero label therefore hides an important distinction between implementation concentration and coordination concentration.

\subsubsection{Multifaceted labels often identify the same core subset}
The high overlap among sociotechnical, technosocial, and superhero categories (0.87--0.94 in Figure~\ref{fig:rq2-heatmap}) should not be read as evidence that the labels are conceptually identical. It reflects how the default ``strong'' threshold works: once a developer is strong on both facets, several aggregated definitions can include them. Thus, the sharpest empirical boundary is between technical and social heroism; the multifaceted categories then describe different ways in which developers cross that boundary.

\subsubsection{Contribution rankings are weakly aligned across facets}
Cross-facet Spearman correlations between technical and social contribution values are weakly negative at the project level (median $\rho \approx -0.11$ across all T$\times$S pairs). This reinforces the set-overlap results: technical and social activity rankings are not simply two views of the same ordering.

\par \textbf{Answer to RQ2:} \textit{Hero identification is comparatively stable within the technical facet and within the social facet, but it changes substantially across facets. Consequently, the chosen measurement facet materially affects who is labelled a hero.}

\subsection{RQ3: Bug-fixing outcomes across hero categories}
Absolute differences in fix and reopen rates are modest, so we interpret RQ3 mainly through severity-wise rankings and their stability rather than through large effect claims.

\begin{figure}[!b]
\centering
\includegraphics[width=\columnwidth]{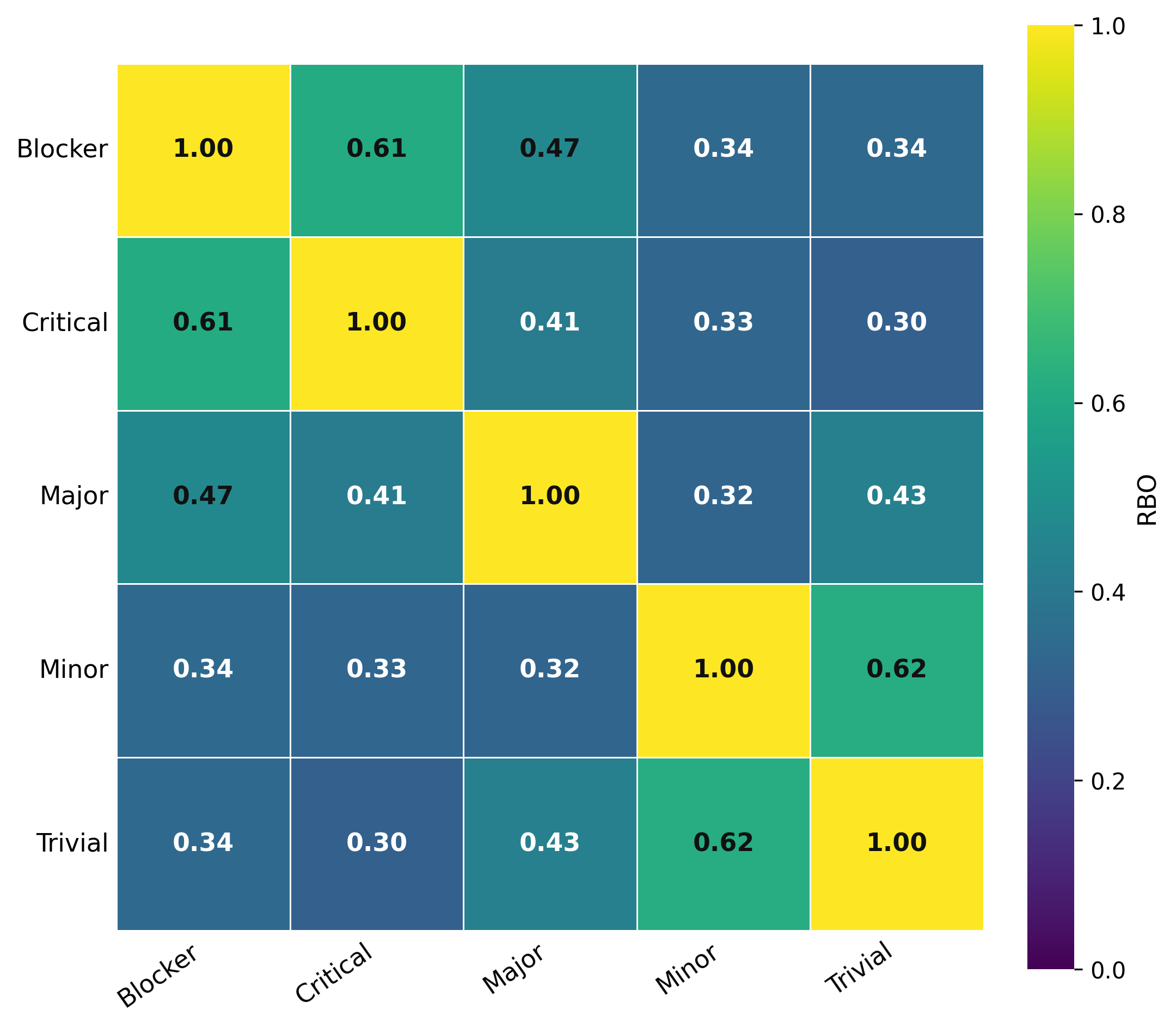}
\caption{RBO ($p{=}0.9$) between severity-wise reopen-rate rankings of aggregated hero types under assignee attribution.}
\Description{Heatmap of pairwise RBO values between blocker, critical, major, minor, and trivial reopen-rate rankings. Blocker--critical and minor--trivial have the highest off-diagonal overlaps, while several cross-severity pairs show lower overlap, indicating that hero-category rankings vary by severity.}
\label{fig:rq3-rbo-heatmap}
\end{figure}

\subsubsection{The best category depends on the outcome}
Table~\ref{tab:rq3-top} shows this divergence: for blocker bugs, \textit{sociotechnical} has the best fix rate, while \textit{technical} has the best reopen rate (tied with technosocial). Fix rate captures whether assigned bugs reach a fixed resolution; reopen rate captures whether fixes remain durable. Thus, the category with the higher fixed proportion is not necessarily the category with the most stable fixes.

\subsubsection{The best category also depends on severity}
Rankings shift across severity levels (Figure~\ref{fig:rq3-rbo-heatmap}). For reopen rate, blocker--critical rankings are more similar (RBO=0.613, $p{=}0.0165$) than blocker--minor rankings (RBO=0.339). This cautions against a severity-agnostic recommendation such as ``assign bugs to the best hero type.'' Different hero categories may be more or less suitable depending on both severity and desired outcome.

\subsubsection{Why the patterns may differ}
One plausible mechanism is that high-severity bugs require both knowledge of the code and coordination with other stakeholders. Sociotechnical developers may be useful for moving such bugs towards resolution because they combine issue-discussion visibility with sufficient technical activity. Lower reopen rates, however, may depend more directly on deep implementation familiarity, which could explain why technical and technosocial categories rank well for reopen rate in severe cases. These interpretations remain hypotheses; our design shows associations, not causal mechanisms.

\subsubsection{Robustness summary}
The main qualitative conclusions are stable across the threshold grid ($\alpha \in \{0.15,0.20,0.25,0.30\}$, $\beta \in \{0.70,0.80,0.85\}$, $q \in \{0.50,0.75\}$). Comparing $q{=}0.50$ with $q{=}0.75$ across the 12 $\alpha$--$\beta$ pairs yields matched aggregated-type prevalence in all 60 cells and matched severity$\times$type$\times$outcome top-category labels in all 600 cells. Tightening the cross-facet threshold reduces multifaceted member rows by 34.9\% on average for sociotechnical and by 6.0\% for technosocial categories. Attribution robustness is also stable: assignee-at-fix-time and combined attribution match the assignee-based reference best-category labels in 50/50 cells.

\begin{table}[!t]
\caption{Best aggregated hero types by severity under assignee attribution. Rates are project-level medians; lower is better for reopen rate and higher for fix rate.}
\label{tab:rq3-top}
\centering
\footnotesize
\setlength{\tabcolsep}{2pt}
\renewcommand{\arraystretch}{1.06}
\begin{tabular*}{\columnwidth}{@{\extracolsep{\fill}}l p{0.40\columnwidth} p{0.40\columnwidth}@{}}
\toprule
\textbf{Severity} & \textbf{Best reopen rate (lower)} & \textbf{Best fix rate (higher)} \\
\midrule
Blocker  &
Technical (2.9, $n{=}63$; tie: technosocial) &
Sociotechnical (93.3, $n{=}60$) \\
Critical &
Technical (3.2, $n{=}68$) &
Technical (84.6, $n{=}69$; tie: technosocial) \\
Major    &
Technical (5.0, $n{=}76$) &
Sociotechnical (85.9, $n{=}71$) \\
Minor    &
Sociotechnical (4.8, $n{=}69$) &
Technical (83.3, $n{=}74$; ties: sociotechnical, technosocial, super) \\
Trivial  &
Technical (0.0, $n{=}56$; ties: social, sociotechnical, technosocial, super) &
Sociotechnical (95.6, $n{=}55$) \\
\bottomrule
\end{tabular*}
\renewcommand{\arraystretch}{1.0}
\end{table}

\par \textbf{Answer to RQ3:} \textit{Fix and reopen differences are modest, but the ranking of hero categories changes by severity and by outcome. Bug-fixing claims about heroes should therefore be severity-aware and checked against alternative thresholds and attribution policies.}

\subsection{Implications}
\subsubsection{Implications for researchers}
Researchers should avoid treating ``hero developer'' as a metric-independent role. The same project can look hero-driven under several measures, while the people identified as heroes differ across facets. Therefore, studies that compare heroes with non-heroes should report at least three pieces of information: the contribution facet used to define heroes, the overlap with plausible alternative definitions, and the sensitivity of conclusions to threshold choices. This changes the takeaway from ``heroes are good'' or ``heroes are risky'' to a more testable claim: \emph{which definition of heroism, measured from which data source, is associated with which outcome?}

\subsubsection{Implications for practitioners}
For project dashboards, staffing, or bug triage, technical and social signals should not be collapsed too early. A top committer and a top issue coordinator may both be important, but they are not interchangeable. For severe bugs, managers may want to distinguish the goal of assignment: fast progress towards a fix may benefit from developers who combine social coordination and technical strength, whereas durable fixes may require deeper technical familiarity. These findings should support triage discussions, not automate personnel judgements.

\subsection{Threats to Validity}
\subsubsection{Construct validity}
Issue priority is an imperfect proxy for severity and may be applied inconsistently across projects. Our technical measures capture volume, breadth, and change magnitude, but they do not directly measure contribution quality, code ownership, or domain expertise. Churn is treated as change magnitude, not as lines written. Our social facet is limited to issue comments and excludes other coordination channels such as code reviews, mailing lists, synchronous meetings, or private discussions. The assignee-based attribution policy is also only an approximation of who fixed a bug; we mitigate this limitation with assignee-at-fix-time and combined attribution variants.

\subsubsection{Internal validity}
Identity linkage across artifacts may be incomplete, potentially splitting one developer into multiple identities. We avoid additional heuristic merging to reduce false joins, but this can lower apparent cross-facet overlap. Our category construction also depends on heuristic thresholds. The Pareto-style $(\alpha,\beta)$ rule and the median-based cross-facet threshold $q$ are practical choices rather than theory-derived cut-offs, although the qualitative picture remains stable across the sensitivity grid. Reopen detection depends on workflow conventions and recorded transitions, so some reopenings may be unobserved.

\subsubsection{External validity}
Our results are based on 77 Apache Software Foundation projects. They may not generalise to ecosystems with different governance models, contribution norms, issue trackers, or review practices.

\subsubsection{Statistical conclusion validity}
Median rate differences across categories are modest and ties are common. We therefore emphasise robustness checks, RBO-based rank stability, and descriptive interpretation rather than strong claims about effect magnitude or causality.

\section{Conclusion}
Heroism in software projects is not only a property of contribution concentration; it is also a consequence of how contribution is measured. In 77 Apache projects, hero projects are common across technical and social measures, but the developers identified as technical heroes and social heroes overlap only modestly. Multifaceted heroism is common and asymmetric: technical heroes often show strong social activity, while social heroes less often show strong technical activity. In bug fixing, category rankings vary by severity and outcome, and fix-rate leaders are not always reopen-rate leaders. These findings support a more careful reporting practice: hero analyses should be facet-aware, severity-aware, and explicit about measurement sensitivity. Future work should incorporate additional coordination channels and develop more theory-grounded role constructions.

\bibliographystyle{ACM-Reference-Format}
\bibliography{references}

\end{document}